\documentstyle[preprint,aps]{revtex}
\tightenlines
\begin{document}
\draft
\title{Entanglement measure for the universal classes of fractons}

\author{Wellington da Cruz and S. Shelly Sharma}
\address{Departamento de F\'{\i}sica,\\
 Universidade Estadual de Londrina, Caixa Postal 6001,\\
Cep 86051-970 Londrina, PR, Brazil\\
E-mail address: wdacruz@exatas.uel.br and shelly@uel.br}
\date{\today}
\maketitle
\begin{abstract}

We introduce the notion of entanglement measure for the universal classes 
of fractons as an entanglement between ocuppation-numbers of fractons in 
the lowest Landau levels and the rest of the many-body system of particles. 
This definition came as an entropy of the probability distribution {\it \`a la} 
Shannon. Fractons are charge-flux systems classified in universal classes of 
particles or quasiparticles labelled by a fractal or Hausdorff dimension 
defined within the interval $1 < h < 2$ and associated with the fractal 
quantum curves of such objects. They carry rational or irrational values 
of spin and the spin-statistics connection takes place in this fractal approach 
to the fractional spin particles. We take into account the fractal von Neumann 
entropy associated with the fractal distribution function which each 
universal class of fractons satisfies. We consider the fractional 
quantum Hall effect-FQHE given that fractons can model Hall states. 
According to our formulation entanglement between occupaton-numbers 
in this context increases with the universality classes of the 
quantum Hall transitions considered as fractal sets of dual 
topological quantum numbers filling factors. We verify that the Hall states have 
stronger entanglement between ocuppation-numbers and so we can consider 
this resource for fracton quantum computing. 

\end{abstract}


\newpage

\section{Introduction}

Entanglement is a foundational characteristic of quantum systems and, more recently, 
has been considered as a physical resource for quantum information processing. There 
is a great effort to find right answers to questions like how to measure and 
classify quantum correlations for implement quantum computation\cite{R1}. 
An entangled state is a state which cannot be written as a direct 
product of states of the parts of a composite quantum system. Several 
entanglement measures are known, for bipartite system 
but no well defined measure exist for mixed multipartite systems. Hence 
a general definition of entanglement measure is an open problem.

In strongly correlated systems such as fractional quantum Hall effect-FQHE, 
these ideas can give us some insight into the understanding 
of this macroscopic complex system. FQHE presents subtle properties of entanglement 
for different Hall states and in the literature is suggested 
that the theory of entanglement developed in the context of quantum 
computing is a suitable tool for this investigation. The relevance 
of this perspective, for example, to a deeper understanding of quantum phase 
transitions, has been emphasized by leading researches. In\cite{R2} 
entanglement properties of the Laughlin wave functions for filling 
factors ( $f=1/m$, with m an odd number ) and for wave functions generated by the 
K-matrix of quantum Hall liquid $( f=2/(2m+1) )$ have been studied considering an 
entanglement measure of indistinguishable fermions. Another one, in terms of 
ocuppation-numbers for fermions, was discussed in\cite{R3}.

In this Letter, we introduce the notion of entanglement measure for the universal 
classes of fractons\footnote{By universal class of fractons we mean a set of 
particles with  rational or irrational values of spin which 
satisfy a specific fractal distribution function in the same 
way that fermions constitute a universal class of particles with semi-integer  
values of spin satisfying the Fermi-Dirac distribution function.}. 
These objects are charge-flux systems which carry 
rational or irrational values of spin and are classified in universal 
classes of particles or quasiparticles labelled by a fractal or Hausdorff 
dimension\footnote{The fractal dimension $h$ can be defined by

\[
h-1=\lim_{R\rightarrow 0}\frac{\ln\;(L/l)}{\ln\;(R)},
\]

 where $L$ is the perimeter of a closed curve, $\Gamma$, and
  $l$ is the usual length for the resolution $R$. The curve is 
 covering with $\frac{l}{R}$ spheres of diameter $R$ and so 
 a fractal curve is scale invariant, 
 self-similar and has a non-integer dimension\cite{R4}.} defined in the interval 
 $1$$\;$$ < $$\;$$h$$\;$$ <$$\;$$ 2$. 
The spin of the particles are related to the Hausdorff dimension by $h=2-2s$, 
$0 < s < \frac{1}{2}$. This expression is a  
physical analogous to the fractal dimension formula 
of the graph of the functions, $\Delta(\Gamma)=2-H$, 
in the context of the fractal geometry, where $H$ is 
known as H\"older exponent, with $0 < H < 1$\cite{R4}. 

The bounds of the fractal dimension $1$$\;$$ < $$\;$$\Delta(\Gamma)$$\;$$ 
<$$\;$$ 2$ need to be obeyed in order for a function 
to be a fractal ( a continuous function but not differentiable\cite{R4} ). 
The bounds of our parameter $h$ are 
defined such that, for $h=1$ we have fermions, for $h=2$ we have bosons, and for 
$1$$\;$$ < $$\;$$h$$\;$$ <$$\;$$ 2$ we have fractons. The H\"older exponent 
characterizes irregular functions which appear in diverse physical systems\cite{R5}. 
The fractal character of the quantum paths has been observed in 
the path integral approach of 
the quantum mechanics\cite{R6}. Thus, 
the fractal dimension is a geometrical parameter 
associated with the quantum curves of fractons. 
Alternatively, the fractal properties of the 
quantum paths can be extracted from the propagators of the particles 
in the momentum space\cite{R7,R9} and so our expression relating 
$h$ and $s$ can once more be justified ( for another view, see below ). 
The physical formula 
introduced by us, when we consider 
the spin-statistics relation $\nu=2s$, is written as $h=2-\nu$, $0 < \nu < 1$. 
In this way, taking into account a mirror symmetry, a fractal spectrum 
has been defined as 

\begin{eqnarray}
\label{e1}
h-1&=&1-\nu,\;\;\;\; 0 < \nu < 1;\;\;\;\;\;\;\;\;h-1=
\nu-1,\;\;\;\;\;\;\; 1 <\nu < 2;\nonumber\\
h-1&=&3-\nu,\;\;\;\; 2 < \nu < 3;\;\;\;\;\;\;\;\;
 h-1=\nu-3,\;
\;\;\;\;\;\; 3 <\nu < 4;\;etc.
\end{eqnarray}

The statistical weight for the universal classes of fractons is given by\cite{R7}

\begin{equation}
\label{e11}
{\cal W}[h,n]=\frac{\left[G+(nG-1)(h-1)\right]!}{[nG]!
\left[G+(nG-1)(h-1)-nG\right]!}
\end{equation}

and from the condition of the entropy be a maximum, we obtain 
the fractal distribution function

\begin{eqnarray}
\label{e.44} 
n[h]=\frac{1}{{\cal{Y}}[\xi]-h}.
\end{eqnarray}

The function ${\cal{Y}}[\xi]$ satisfies the equation 

\begin{eqnarray}
\label{e.4} 
\xi=\biggl\{{\cal{Y}}[\xi]-1\biggr\}^{h-1}
\biggl\{{\cal{Y}}[\xi]-2\biggr\}^{2-h},
\end{eqnarray}

\noindent with $\xi=\exp\left\{(\epsilon-\mu)/KT\right\}$. 
The statistical weight can be written in terms of gamma function

\begin{eqnarray}
\label{e111}
{\cal W}[x,y]&=&\frac{\Gamma(x+y+1)}{\Gamma(x+1)\Gamma(y+1)},
\end{eqnarray}

\noindent where $x[h]=N=nG$ and $y[h]=G+(N-1)h+1-2N$, such that for bosons $y[h=2]=G-1$ 
and for fermions $y[h=1]=G-N$. For particles with spin defined in the interval 
$0\leq s \leq \frac{1}{2}$, we obtain $y[s]=G-(N-1)2s-1$, so in the large 
$N$ limit we have $h=2-2s$ and for the statistical parameter within the interval
$0\leq \nu \leq 1$, we obtain $y[\nu]=G-(N-1)\nu-1$, and again in the large $N$ limit 
$h=2-\nu$, so $\nu=2s$, i.e. the spin-statistics connection is established. 
We can check that all the interpolating expressions have the same bounds. 
Each expression reduces to the other one. 
  
We understand the 
fractal distribution function as a quantum-geometrical 
description of the statistical laws of nature, 
since the quantum path is a fractal curve ( a point noted by Feynman ) and this 
reflects the Heisenberg uncertainty principle. The Eq.(\ref{e.44}) 
embodies nicely this subtle information about the quantum paths 
associated with the particles. 

We can obtain for any class its distribution function considering the
Eqs.(\ref{e.44},\ref{e.4}). For example, 
the universal class $h=\frac{3}{2}$ with distinct values of spin 
$\biggl\{\frac{1}{4},\frac{3}{4},\frac{5}{4},\cdots\biggr\}_{h=\frac{3}{2}}$, 
has a specific fractal distribution

\begin{eqnarray}
n\left[\frac{3}{2}\right]=\frac{1}{\sqrt{\frac{1}{4}+\xi^2}}.
\end{eqnarray}

\noindent This result coincides with another 
one of the literature of fractional spin particles for 
the statistical parameter $\nu=\frac{1}{2}$\cite{R10}, 
however our interpretation is completely distinct. This particular example, 
shows us that the fractal distribution is the same 
for all the particles into the universal class labelled by $h$ and with 
different values of spin ( consider the fractal spectrum, for a simple check ). 
Thus, we emphasize that in our formulation the spin-statistics 
connection is valid for such fractons. The authors in\cite{R10} 
ignored this possibility. Therefore, 
our results give another perspective for the fractional 
spin particles or anyons\cite{R11}. By the way, we can obtain straightforward the 
Hausdorff dimension associated to the quantum paths of 
the particles with any value of spin. This constitutes a fine result of our approach.  

We also have
 
\begin{eqnarray}
\xi^{-1}=\biggl\{\Theta[{\cal{Y}}]\biggr\}^{h-2}-
\biggl\{\Theta[{\cal{Y}}]\biggr\}^{h-1},
\end{eqnarray}

\noindent where

\begin{eqnarray}
\Theta[{\cal{Y}}]=
\frac{{\cal{Y}}[\xi]-2}{{\cal{Y}}[\xi]-1}
\end{eqnarray}

\noindent is the single-particle partition function. 
We verify that the classes $h$ satisfy a duality symmetry defined by 
${\tilde{h}}=3-h$. So, fermions and bosons come as dual particles. 
As a consequence, we extract a fractal 
supersymmetry which defines pairs of particles $\left(s,s+\frac{1}{2}\right)$. 
This way, the fractal distribution function appears as 
a natural generalization of the fermionic and bosonic 
distributions for particles with braiding properties. Therefore, 
our approach is a unified formulation 
in terms of the statistics which each universal class of 
particles satisfies, from a unique expression 
we can take out any distribution function.

The fractal von Neumann entropy per state in terms of the 
average occupation number is given as\cite{R7,R8} 

\begin{eqnarray}
\label{e5}
{\cal{S}}_{G}[h,n]&=& K\left[\left[1+(h-1)n\right]\ln\left\{\frac{1+(h-1)n}{n}\right\}
-\left[1+(h-2)n\right]\ln\left\{\frac{1+(h-2)n}{n}\right\}\right]
\end{eqnarray}

\noindent and it is associated with the fractal distribution function Eq.(\ref{e.44}).

In\cite{R8} we have considered, a microstate probability as

\begin{equation}
\label{e12}
{\cal P}(n)=p^{nG}q^{[n(h-2)+1]G-(h-1)},
\end{equation}

\noindent where $nG=N$ is the number of particles, $G$ is the number of states and 
$p+q=1$. Also we have that the total propability is unity 

\begin{equation}
\sum_{n}{\cal W}(n){\cal P}(n)=1,
\end{equation}

\noindent and differentiating this expression with respect to p, we find

\begin{equation}
n\left[\frac{q}{p}+2-h\right]G=G-(h-1),
\end{equation}

\noindent and as $G >> (h-1)$ $( 1 < h < 2 )$, in the large $G$ limit, 
we recover the Eq.(\ref{e.44}), with the definition of the function 
$\left[\frac{q}{p}+2\right]\equiv\cal{Y}[\xi]$, which satisfies the Eq.(\ref{e.4}).

The microstate probabilities are written as $p=\frac{1}{{\cal{Y}}[\xi]-1}$ and 
$q=\frac{{\cal{Y}}[\xi]-2}{{\cal{Y}}[\xi]-1}$, and the Eq.(\ref{e.44}) 
as $n[h]=\frac{p}{q+(2-h)p}$. On the other hand, the von Neumann entropy 
in terms of the matrix density is found to be\cite{R12}

\begin{eqnarray}
\label{e10} 
\frac{{\cal S}}{K}&=&-{\rm Tr}\;\rho\ln\rho\\
&=&-\sum_{n}{\cal W}(n){\cal P}(n)\ln{\cal P}(n),\\
\nonumber
\end{eqnarray}

\noindent and considering our previous definitions we reobtain 
the fractal von Neumann entropy as

\begin{equation}
\label{e5}
{\cal{S}}_{G}[h]=n\;K\biggl[({\cal{Y}}[\xi]-1)\ln({\cal{Y}}[\xi]-1)-
({\cal{Y}}[\xi]-2)\ln({\cal{Y}}[\xi]-2)\biggr].
\end{equation}

\noindent Finally, in terms of $p$ and $q$, we get

\begin{equation}
\frac{{\cal{S}}_{G}[h]}{K}=\frac{1}{q+(2-h)p}\left\{-p\ln p - q\ln q\right\}.
\end{equation}

\noindent Now, we define an entanglement measure\footnote
{Or entanglement fractal von Neumann entropy.} for 
the universal classes of fractons in terms of the 
probability distribution $p$ as:

\begin{equation}
\label{e.43}
{\cal{E}}[h,p]=\frac{1}{1-(h-1)p}\left\{-p\log_{2}p -(1-p)\log_{2}(1-p)\right\},
\end{equation}

\noindent where $0 \leq p\leq 1$, is the probability of the  system to 
be in a microstate with 
entanglement between ocuppation-numbers of the modes 
considered empty, partially or completely filled. Thus, 
an entangled state of fermions, for example, with one particle and three modes is written as

\[
|\Phi>=c_{1}|110>+c_{2}|101>+c_{3}|011>,
\]

\noindent with $p=|c_{i}|^2$ and $\sum_{i}|c_{i}|^2=1$. For fractons, for example of    
the class $h=\frac{3}{2}$, with $3$ modes 
and $4$ particles\footnote{In particular, we have here the maximum of 
two particles for each mode, so 
the entanglement between the $3$ modes presents this configuration.} we have the configuration

\[
|\Psi>=c_{1}|121>+c_{2}|022>+c_{3}|211>+c_{4}|202>+c_{5}|112>+c_{6}|220>,
\]

\noindent and the amount of entanglement is given by:
  
\begin{eqnarray}
\label{e.130}
{\cal{E}}\left[h=\frac{3}{2},p\right]&=&
\frac{2}{2-p}\left\{ -p\log_{2}p - (1-p)\log_{2}(1-p)\right\}\\
&=&\sum_{i=1}^6\left\{\frac{2}{2-|c_{i}|^2}\left[-|c_{i}|^2\log_{2}|c_{i}|^2-
(1-|c_{i}|^2)\log_{2}(1-|c_{i}|^2)\right]\right\}.
\end{eqnarray}

\noindent The Eq.(\ref{e.43}) for fermions reduces to

\begin{eqnarray}
\label{e.21}
{\cal{E}}[h=1,p]&=&-p\log_{2}p-(1-p)\log_{2}(1-p),
\end{eqnarray}

\noindent and the amount of entanglement for the entangled state 
considered above is given by

\begin{eqnarray}
\label{e.131}
{\cal{E}}[h=1,p]&=&\sum_{i=1}^3\left\{-|c_{i}|^2\log_{2}|c_{i}|^2-
(1-|c_{i}|^2)\log_{2}(1-|c_{i}|^2)\right\}.
\end{eqnarray}

\noindent The expression Eq.(\ref{e.21}) coincides with another 
one in Ref.\cite{R3} for the entanglement measure between ocuppation-numbers 
of different single particle basis states in the context of the FQHE. On the other hand, 
in Ref.\cite{R2} an entanglement measure of indistinguishable fermions 
was considered taking into account the Laughlin wave functions and those ones generated by the 
K-matrix of the quantum Hall liquid. Therefore, our approach consider 
different systems of particles ( fractons ), i.e., charge-flux systems 
which carry rational or irrational values of spin, so in this way we obtain results in agreement 
with those ones reported in Refs.\cite{R2,R3}. 
These points we will discuss in the next section.

\section{Entanglement in the FQHE}

In\cite{R7} we have considered a fractal approach to the FQHE\cite{R13} 
with Hall states modeled by fractons. According to our formulation the quantum Hall 
state associated with a specific filling factor is a fracton 
state with value of spin $s=\nu/2$. The filling factor which characterizes 
the quantization of the Hall resistance, has the same value of the 
statistical parameter, i.e. $\nu=f ( numerically ) $, where $f$ is defined 
by $f=N\frac{\phi_{0}}{\phi}$, and 
$N$ is the electron number, 
$\phi_{0}$ is the quantum unit of flux and
$\phi$ is the flux of the external magnetic field throughout the sample. 
The spin-statistics relation is given by 
$\nu=2s=2\frac{\phi\prime}{\phi_{0}}$, where 
$\phi\prime$  is the flux associated with the charge-flux 
system which defines the fracton $(h,\nu)$. In this way the universality 
classes of the quantum Hall transitions satisfy some properties 
of a subgroup of the modular group $SL(2,{\bf Z})$ related with the 
Farey sequences of rational numbers. The 
transitions allowed are those generated by the condition
 $\mid p_{2}q_{1}
-p_{1}q_{2}\mid=1$, 
with $\nu_{1}=\frac{p_{1}}{q_{1}}$ and $\nu_{2}=\frac{p_{2}}{q_{2}}$\cite{R14,R7}. 
This way, we define the universality classes of the quantum Hall 
transitions in terms of fractal sets labelled by the Hausdorff dimension. 
We verify that the filling factors 
experimentally observed  appear into the classes $h$ and from the definition of duality 
between the fractal sets, we note that the FQHE occurs in pairs 
of dual filling factors. These quantum numbers get 
their topological character from the 
fractal dimension associated with the quantum paths. Our results show clearly 
which the FQHE has a fractal-like structure and this deeper feature is 
revealed by robust mathematical concepts. Another 
fractal formulation to the FQHE was discussed in\cite{R15} 
and considered as an approach to be explored for 
understanding the subtle properties of the FQHE\cite{R16}. However, we observe that our 
program anticipated this suggestion just considering properly 
ideas of the fractal geometry\cite{R7,R8}.

The entanglement properties of fractons ( Hall states ) 
can be now analyzed considering the Eq.(\ref{e.43}). For distinct classes of 
fractons we have verified that ${\cal{E}}[h,p]={\cal{E}}[h,1-p]$. 
We can check for the classes of particles $h=1,\frac{4}{3},
\frac{3}{2},\frac{5}{3}$, via the graphic ${\cal{E}}\times p$, $0\leq p \leq 1$, 
which the entanglement fractal von Neumann entropy is a concave function 
and increases in the interval $ 1 < h < 2 $, for instance, 
${\cal{E}}[h=4/3]< {\cal{E}}[h=3/2]<
{\cal{E}}[h=5/3]$. Consider now, the sequence  

\begin{eqnarray}
&&\cdots\rightarrow\biggl\{\frac{2}{3},\frac{4}{3},\frac{8}{3},
\cdots\biggr\}_{h=\frac{4}{3}}\rightarrow\;\;
\biggl\{\frac{1}{2},\frac{3}{2},\frac{5}{2},
\cdots\biggr\}_{h=\frac{3}{2}}\rightarrow
\biggl\{\frac{1}{3},\frac{5}{3},\frac{7}{3},
\cdots\biggr\}_{h=\frac{5}{3}}\rightarrow\cdots,
\end{eqnarray}

\noindent and for the entanglement measure written in 
terms of the filling factors

\begin{eqnarray}
\label{e12}
{\cal{E}}[2-\nu,p]&=&\frac{1}{1-(1-\nu)p}\left\{-p\log_{2}p -(1-p)\log_{2}(1-p)\right\}, 
\end{eqnarray}

\noindent with $0 < \nu < 1$, and we confirme that ${\cal{E}}[\nu=2/3]< {\cal{E}}[\nu=1/2]<
{\cal{E}}[\nu=1/3]$. For the other members of the classes we need to consider the 
fractal spectrum Eq.(\ref{e1}). This way, we verify that the Eq.(\ref{e.43}) 
for the class $h$, is 
the same for all the members of the class and so, in terms of their entanglement content, 
different Hall states are equivalent. The understanding that something in this sense 
can be provided by a quantitative theory of entanglement for 
complex quantum systems was envisaged by Osborne-Nielsen 
in\cite{R17}. Therefore, we have obtained  
a result, in the context of the FQHE, which just realizes this perception. {\it Observe that our approach 
gives information about the entanglement for any possible wave function 
associated with a specific value of the filling factor}. In another route we can consider 
the LLL for fractons, i.e. if the temperature 
 is sufficiently low and $\epsilon <\mu$, we can check 
 that the mean ocuppation number Eq.(\ref{e.44}) is given 
 by $n=\frac{1}{2-h}$, and so the fractal parameter $h$ 
 regulates the number of particles in each quantum state. For 
 $h=1$,$\;$$n=1$;$\;$$h=2$,$\;$$n=\infty$;$\;$ etc. At $T=0$ and $\epsilon> \mu$, $n=0$
 if $\epsilon > \epsilon_{F}$ and $n=\frac{1}{2-h}$ 
 if $\epsilon < \epsilon_{F}$, hence we get a step distribution, 
 taking into account the Fermi energy $\epsilon_{F}$ and $h\neq 2$. We can check that for 
 $h=\frac{4}{3},\frac{3}{2},\frac{5}{3}$ we obtain $n=\frac{3}{2},\frac{2}{1},\frac{3}{1}$, 
 respectively.  In the first case we have three particles for two states, in the second 
 case two particles for one state and in the last case three particles for one state. So when 
 we run in the interval $ 1 < h < 2$ we gain more particles for each possible state. In 
 some sense fractons can be understood as {\it quasifermions} when near 
 the universal class $h=1$ and as $quasibosons$ when near the universal class $h=2$. 
 The entanglement of the FQHE increases because we have more particles ( fractons ) 
 and less states. On the other hand, in terms of the filling factors, 
 the average ocuppation number can be written as
 $n=\frac{1}{\nu}$, $0 < \nu < 1$;  $n=\frac{1}{2-\nu}$, $1 < \nu < 2$; $n=\frac{1}{\nu-2}$, 
 $2 < \nu < 3$; etc. 
 We obtain the pairs $\left(\nu=\frac{2}{3}, n=\frac{3}{2}\right)$; $\left(\nu=\frac{1}{2}, 
 n=\frac{2}{1}\right)$;
 $\left(\nu=\frac{1}{3}, n=\frac{3}{1}\right)$; $\left(\nu=\frac{4}{3}, n=\frac{3}{2}\right)$;
$\left(\nu=\frac{3}{2}, n=\frac{2}{1}\right)$; $\left(\nu=\frac{5}{3}, n=\frac{3}{1}\right)$; 
$\left(\nu=\frac{8}{3}, n=\frac{3}{2}\right)$;
$\left(\nu=\frac{5}{2}, n=\frac{2}{1}\right)$; $\left(\nu=\frac{7}{3}, n=\frac{3}{1}\right)$; 
etc. The behaviour of the step distribution confirmes 
our former analysis: the ground state of the FQHE is a stronger entangled state and the 
entanglement between ocuppation-numbers of fractons in the LLL and the rest of 
the system shows us quantum correlations which can be quantified.

All these results agree with the entanglement properties of the Laughlin 
wave functions and those generated by the K-matrix\cite{R2}. On the other hand, 
the FQHE understood in terms of the composite fermions or composite 
bosons are non-entangled as observed in\cite{R3}, so in contrast, fractons appear 
as a suitable system for study the quantum correlations of the FQHE. 
Thus the suggestion that ideas of the quantum information science can give insights for 
understanding some complex quantum systems\cite{R17} is manifested in 
our definition of entanglement measure for the universal classes of fractons. 
The universality classes of the quantum Hall transitions as fractal sets of dual 
topological quantum numbers filling factors, according to our formulation, have increasing 
entanglement in the interval $ 1 < h < 2 $ and this suggests fracton 
qubits as a physical resource for quantum computing. In the literature, FQHE qubits 
associated with the geometrical characteristic of the fractional spin particles 
have been exploited\cite{R18}. The quantum Hall phase transitions discussed by us 
were obtained considering global properties as the modular symmetry and 
the Hausdorff dimension associated to the quantum paths of the particles, 
so some peculiarities of the FQHE, in particular, do not depend on the dynamical 
aspects or other details of this strongly interacting system\cite{R7,R8}.

\section{Conclusions}

The discussion of the FQHE in terms of fractons shows us the 
potential application of this physical support for the implementation 
of quantum computing. The topological character of these objects 
is crucial against problems of the decoherence. Entangled fracton states 
can be considered as a stable resource for a topological 
quantum computation, i.e., a fault-tolerant quantum computation. We observe 
again that fractons carry rational or irrational values of spin and 
they obey the spin-statistics connection. 
In the literature geometric phases and the concept of anyons have been 
explored for implement quantum gates because they are 
robust against random noise of the environment\cite{R19}.

Finally, we have introduced the notion of entanglement measure for the universal 
classes of fractons, where concepts of the fractal geometry appear naturally 
and gives us the opportunity to extract information about 
quantum correlations of the ground state of the FQHE. We can obtain the entanglement properties 
of any possible wave function associated with 
a specific value of the filling factor such as: Laughlin wave functions\cite{R13}, 
wave functions generated by the K-matrix of quantum Hall liquid\cite{R20}, the Pfaffian 
trial wave function\cite{R21} associated with nonabelian quantum statistics, 
Jain functions\cite{R22}, Halperin functions\cite{R23}, etc. This way, in some sense, 
we have elaborated a unifying framework for understanding subtle 
properties of complex quantum systems. Our approach reveals us the fractal nature 
of this FQHE-phenomenon\cite{R7,R8}. The possibility of to 
establish a bridge between quantum information theory 
and the quantum Hall transitions goes to the direction 
of some ideas of research in the literature 
and considered of extreme importance\cite{R17}. As we saw, 
our results agree with other ones\cite{R2,R3} and open an avenue 
for we speculate on a fracton quantum computing.

\end{document}